\documentclass[12pt]{article}
\usepackage{epsf,amsfonts,amssymb,epsfig,amsmath,mathtools,graphics,slashed}
\usepackage{hyperref,hep,graphicx,subfig}
\usepackage{rotating}
\usepackage{xcolor}
\allowdisplaybreaks

\definecolor{greenish}{RGB}{0,190,0}

\definecolor{yellowish}{RGB}{190,190,0}

\newcommand{\nn}{\notag \\}

\makeatletter

\@addtoreset{equation}{section}
\makeatother

\begin{document}

\begin{titlepage}

\vfill

\begin{flushright}
DCPT-19/13
\end{flushright}

\vfill

\begin{center}
   \baselineskip=16pt
   {\Large\bf Hydrodynamics of broken global symmetries in the bulk}
  \vskip 1.5cm
  \vskip 1.5cm
      Aristomenis Donos$^1$, Daniel Martin$^1$, Christiana Pantelidou$^1$ and Vaios Ziogas$^2$\\
%   \vskip .6cm
%      \begin{small}
%      \textit{$^1$Centre for Particle Theory and Department of Mathematical Sciences,\\ Durham University,
%       Durham, DH1 3LE, U.K.}
%        \end{small}\\   
    \vskip .6cm
      \begin{small}
      \textit{$^1$ Centre for Particle Theory and Department of Mathematical Sciences,\\ Durham University,
       Durham, DH1 3LE, U.K.}
        \end{small}\\
      \vskip .6cm
      \begin{small}
      \textit{$^2$Shanghai Center for Complex Physics, School of Physics and Astronomy,\\
        Shanghai Jiao Tong University, Shanghai 200240, China}
        \end{small}\\
         
\end{center}

\vfill

\begin{center}
\textbf{Abstract}
\end{center}
\begin{quote}
We consider holographic theories at finite temperature in which a continuous global symmetry in the bulk is spontaneously broken. We study the linear response of operators in a regime which is dual to time dependent, long wavelength deformations of solutions generated by the symmetry. By computing the boundary theory retarded Green's function we show the existence of a gapless mode with a diffusive dispersion relation. The diffusive character of the mode is compatible with the absence of a conserved charge from the field theory point of view. We give an analytic expression for the corresponding diffusion constant in terms of thermodynamic data and a new transport coefficient $\sigma_{b}$ which is fixed by the black hole horizon data. After adding a perturbative source on the boundary, we compute the resulting gap $\delta\omega_{g}$ as a simple function of $\sigma_{b}$ and of data of the thermal state.
\end{quote}

\vfill

\end{titlepage}

\setcounter{equation}{0}

\section{Introduction}

The AdS/CFT correspondence is a powerful duality that allows us to make precise computations within a large class of strongly coupled conformal field theories. With applications to real systems in mind, the biggest strength of the duality is that it allows us to perform computations in, and ultimately characterise, systems at finite temperature and chemical potential from a microscopic point of view.

One of the milestone developments in the dawn of AdS/CMT \cite{Herzog:2009xv,Hartnoll:2009sz,Hartnoll:2016apf} was the observation that boundary symmetry breaking may be realised in the bulk in a very natural way \cite{Gubser:2008px,Hartnoll:2008kx}. In this framework, a bulk field develops a perturbative instability below a critical temperature $T_{c}$. The resulting zero mode yields a new branch of back-reacted black hole spacetimes in which one or more symmetries can be broken in the bulk. By lowering the temperature of these broken phase black holes down to zero temperature, one can realise new ground states emerging from the cold event horizon \cite{Horowitz:2009ij,Gubser:2009cg}.

When a continuous bulk symmetry is broken, acting with the symmetry will generate new bulk solutions with the same free energy. If the symmetry is gauged, the component that transforms the VEVs of boundary operators is a global boundary symmetry, and has a Noether current which extends into the bulk as the gauge field. On the other hand, in the case of a global internal bulk symmetry\footnote{Global symmetries are expected to be broken in a quantum theory of gravity \cite{Banks:2010zn,Harlow:2018tng} Nevertheless, they are perfectly well-behaved in the classical low-energy limit.}, there is no longer a conserved charge. Nevertheless, such a symmetry will still transform the VEVs, leading to inequivalent solutions with the same free energy.

As a result, the thermodynamic susceptibilities of operators whose VEVs transform non-trivially under global bulk symmetries will be infinite, signalling long-ranged correlations at finite temperature. In this paper we study the linear response of such operators via their retarded Green's function, in a hydrodynamic limit\footnote{Among other results, the authors of \cite{Anninos:2010sq} considered the hydrodynamic limit of Green's functions in the case of gauged symmetry breaking in the bulk up to linear order in the wavenumber giving a holographic calculation of the speed of sound.} of long wavelengths and small frequencies. In this limit, we give analytic expressions in terms of black hole horizon data and thermodynamic quantities.

From the pole structure of our answer for the Green's function, we extract the hydrodynamic mode associated with the breaking of a global $U(1)$ symmetry in the bulk. As we will explicitly show, the mode is diffusive and we determine the diffusion constant in terms of thermodynamic quantities and a new transport coefficient $\sigma_{b}$ which is fixed by black hole horizon data. This is in contrast to bulk gauge symmetries, in which the Ward identity fixes a linear dispersion relation at leading order in the wavenumber, with the speed of sound given by the charge and current susceptibilities.

When the mode that we study in this paper mixes with the thermoelectric transport currents, it plays a crucial role in the physics of pinning of density waves. This happens when the VEV of the condensed operator in the density wave phase breaks translations, which causes the gapless excitations to couple to the heat current and, if electric charge is present, the electric current. Depending on whether translations are explicitly broken or not one has two cases to consider.

When an explicit background lattice is present, as discussed in \cite{Donos:2019tmo}, the mode we are examining in this paper will couple to the thermoelectric diffusive modes and enlarge the hydrodynamic description of the system by one more degree of freedom \cite{Donos:2019hpp}. The phenomenology becomes more interesting after adding the effects of pinning by explicitly breaking the bulk $U(1)$ symmetry through perturbative boundary sources. At frequencies much smaller that the infinitesimally small gap, the density wave related to the complex order parameter is pinned and it doesn't contribute to transport. However, as soon as one reaches frequencies close to the gap, the density wave gets excited, contributing to the thermoelectric transport properties.

When translations are not explicitly broken by a lattice, the longitudinal excitations of the normal phase consist of a sound mode and an incoherent mode which is diffusive. It would be interesting to know the role of the gapless mode we study here in setups where translations are not explicitly broken. Once again, it can mix with the conserved momentum currents \cite{Amoretti:2019cef} producing sound modes this time. We expect the gapless mode to mix with the incoherent diffusive mode \cite{Davison:2015taa}, via its overlap with the heat current.

Having stressed its significance in the physics of density waves, here we shall focus on the relevant gapless mode when it does not mix with the thermoelectric transport currents. We will achieve this by considering thermodynamic backgrounds in which the VEV of the condensed complex operator does not break translations. Moreover, we will study the effects of perturbatively deforming the theory to explicitly break the internal bulk $U(1)$ symmetry, thereby introducing a gap in the dispersion relation of the aforementioned mode. By performing a systematic perturbative expansion, we compute the gap and express it as a function of the VEV of the condensed operator, the thermodynamic susceptibility of the deformed theory, and the transport coefficient $\sigma_{b}$. In the case where the order parameter breaks translations, this small explicit breaking deformation plays the role of a pinning parameter against an emerging sliding mode. In the setup of \cite{Donos:2019tmo}, this relaxation time is independent of the momentum relaxation time due to the presence of a holographic lattice. However, in the setups presented in \cite{Alberte:2017oqx,Alberte:2017cch,Amoretti:2018tzw}, it also sets the relaxation time of momentum dominating the transport properties of primarily heat in the system.

The paper is organised as follows. In section \ref{sec:setup} we give an introduction to the class of holographic models we are considering where the scenario of our paper is realised. Our discussion includes some general remarks regarding holographic renormalization of our holographic models as well as general statements on the thermodynamics of the black hole solutions we will be interested in. Section \ref{sec:diffusion} contains the first of our main results. We study the retarded Green's function of the operators generated by the action of the bulk symmetry and derive the dispersion relation of the diffusive pole. In section \ref{sec:gap} we give the mode a gap by performing a symmetry breaking deformation in a perturbative manner. The formula for the gap is the second main result of our paper. We have performed a number of numerical checks of our analytic formulae for the Green's function and the gap in section \ref{sec:numerics} where we present a simple model that realises our setup of section \ref{sec:setup}. We summarise in section \ref{sec:discussion} with a discussion of our results.

\textbf{Note added:} While finalising our paper, v2 of \cite{Amoretti:2018tzw} appeared on the arxiv containing a new discussion on the diffusive mode we identify in our section \ref{sec:diffusion}. The authors also discuss the gap which we study in our section \ref{sec:gap}, where we express it in terms of dual field theory quantities.

\section{Set-up}\label{sec:setup}
In order to realise the desired holographic phase transition, we will consider a $d+1$ dimensional gravitational theory which contains at least the bulk metric and a complex scalar $Z$. The bulk part of the action is of the general form
\begin{align}\label{eq:action_general}
S_{bulk}=\int d^{d+1}x\,\sqrt{-G}\left(\frac{1}{2}W(G_{\mu\nu}, \partial G_{\mu\nu},\partial^{2} G_{\mu\nu},\cdots,Z\bar{Z},\mathcal{F})-\frac{1}{2}Q(Z\bar{Z},\mathcal{F})\,\partial_{\mu}Z\,\partial^{\mu}\bar{Z}\right)\,,
\end{align}
where the functions $W$ and $Q$, apart from the metric $G_{\mu\nu}$ and its derivatives, can depend on a number of fields and their derivatives which we are collectively calling $\mathcal{F}$. In order for the kinetic term of the complex scalar to be well defined, we assume that $Q$ is a positive function and that $Q(0,0)=1$. The dependence of $W$ and $Q$ on the complex scalar $Z$ is only through its modulus $u\equiv Z\bar{Z}$ and so the theory possesses a global $U(1)$ symmetry which simply rotates $Z$ in the complex plane. The second basic requirement is that $W$ is such that the equations of motion resulting from \eqref{eq:action_general} allow us to set $\mathcal{F}=0$ and also obtain $AdS_{d+1}$ with unit radius as a solution
\begin{align}\label{eq:ads_sol}
ds^{2}=r^{2}\,\left(-dt^{2}+\delta_{ij}dx^{i}dx^{j}\right)+\frac{dr^{2}}{r^{2}},\quad \mathcal{F}=0,\quad Z=0\,.
\end{align}
The fluctuations of $Z$ around the vacuum \eqref{eq:ads_sol} satisfy
\begin{align}
\nabla^{2}\delta Z-m^{2}\,\delta Z=0,\quad m^{2}=-\left. \partial_{u}W\right|_{\mathcal{F}=0,\,Z=0}\,,
\end{align}
yielding an asymptotic expansion of the form
\begin{align}
\delta Z(t,x^{i},r)=\delta z_{s}(t,x^{i})\,r^{\Delta -d}+\cdots +\delta z_{v}(t,x^{i})\,r^{-\Delta}+\cdots\,,
\end{align}
with $\Delta=\frac{d}{2}+\sqrt{\frac{d^{2}}{4}+m^{2}}$. The complex constants of integration $\delta z_{s}$ and $\delta z_{v}$ play the role of source and VEV of a complex operator on the field theory side with dimension $\Delta$.

For our theory \eqref{eq:action_general}, it is consistent to write down an ansatz of the form
\begin{align}\label{eq:bh_sol}
ds^{2}&=-U(r)\,dt^{2}+\frac{dr^{2}}{U(r)}+g_{ij}(r)\,dx^{i}dx^{j}\,,\nn
Z&=\rho(r)\,e^{i\,k_{j}x^{j}}\,,
\end{align}
and the bulk fields $\mathcal{F}=\mathcal{F}_{0}$ to be consistent with the ansatz \eqref{eq:bh_sol}. For $k_{j}\neq0$, the background solution \eqref{eq:bh_sol} is periodic and preserves a diagonal combination of the global $U(1)$ symmetry and translations in the direction parallel to $k_{j}$, while breaking the orthogonal combination. This breaking can be either explicit or spontaneous, depending on whether we have a source for $\rho$ or just a VEV. When $k_{j}=0$, only the global $U(1)$ symmetry is broken.

For the background solutions we would like to consider thermal states in which the operator dual to $Z$ has taken a VEV spontaneously. For this reason we impose the asymptotic expansion,
\begin{align}\label{eq:ads_as}
U(r)&=r^{2}+\mathcal{O}(r^{0})\,,\nn
g_{ij}(r)&=r^{2}\,\delta_{ij}+\mathcal{O}(r^{0})\,,\nn
\rho(r)&=\frac{\rho_{v}}{r^{\Delta}}+\cdots\,,
\end{align}
close to the conformal boundary of $AdS_{d+1}$. We also impose regularity near the black hole horizon at $r=r_{h}$ by demanding 
\begin{align}\label{eq:hor_as}
U(r)&=4\pi T \left(r-r_{h}\right)+\mathcal{O}\left((r-r_{h})^{2}\right)\,,\nn
g_{ij}(r)&=g^{h}_{ij} +\mathcal{O}\left((r-r_{h})\right)\,,\nn
\rho(r)&=\rho_{h} +\mathcal{O}\left((r-r_{h})\right)\,,
\end{align}
along with appropriate regularity conditions on the rest of the fields $\mathcal{F}$. Here $T$ is the Hawking temperature of the black hole, which is identified with the temperature of the dual field theory.

\subsection{Holographic renormalization and thermodynamics}\label{subsec:thermo}

Thermodynamics is going to play a key role in expressing our main results in terms of physical quantities of the dual field theory. The free energy density is equal to the on-shell Euclidean action $w=\frac{T}{V}I_{E}$, where $V$ is the volume of the unit cell and $I_{E}=-i\,S$ is the Euclidean version of the regulated on-shell action
\begin{align}\label{eq:action_total}
S= S_{bulk}+S_{ct}\,.
\end{align}

The counterterm $S_{ct}$ should be expressed as a local, invariant functional of the induced fields on a constant-$r$ hypersurface \cite{Skenderis:2002wp}. Writing a precise form of $S_{ct}$ for the full theory can be a quite complicated task and it depends on the details of the bulk action $S_{bulk}$. However, in the regimes which will interest us, we will manage to extract a lot of information without knowing the precise details.

Close to the conformal boundary, the bulk function $Z$ will in general admit the expansion
\begin{align}
Z=\frac{z_{s}}{r^{d-\Delta}}+\cdots+\frac{z_{v}}{r^{\Delta}}+\cdots\,.
\end{align}
As far as the background is concerned, we will first consider the one point function of the dual field theory operator $\mathcal{O}_{Z}$ as well as the variation of $w$ with respect to $k_{j}$, in situations where $|z_{s}|\ll |z_{_{v}}|$. In the main part of the paper we will focus on the hydrodynamic limit of the retarded Green's functions by considering perturbations around the background. This will introduce a time dependent perturbative source $\delta z_{s}$ with bulk response $\delta z_{v}$. As we will explain momentarily, in this limit we will be in a similar situation, where $|\delta z_{s}|\ll |\delta z_{_{v}}|$.

After introducing the deformations we wish to consider and under the mild assumption that $Q\to1$ at the boundary, the on-shell variation $\delta S_{bulk}$ of \eqref{eq:action_general} gives only one finite boundary term proportional to $\frac{1}{2}(\Delta-d)\,(\bar{z}_{s}\,\delta z_{v}+z_{s}\,\delta \bar{z}_{v})$. As usual, the counterterm action $S_{ct}$ is required to make the bulk variational problem well defined \cite{Papadimitriou:2005ii}  by cancelling this term in a variation. Moreover, $S_{ct}$ should not introduce new finite terms proportional to $\delta z_{v}$. This consideration suggests that the only finite term in $S_{ct}$ containing $z_{v}$ should result in $\frac{1}{2}(d-\Delta)\,(\bar{z}_{s}\, z_{v}+z_{s}\,\bar{z}_{v})$. However, this argument does not fix the finite terms in $S_{ct}$ which do not depend on $z_{v}$. These can in general depend on the sources $z_{s}$, as well as the asymptotic data of the metric and the other fields $\mathcal{F}_{as}$ in our bulk theory. In summary, the finite part of the counterterms will take the form
\begin{align}\label{eq:Sct_fin}
S_{ct,fin}=\int d^{d}x\,\frac{1}{2}(d-\Delta)\,(\bar{z}_{s}\, z_{v}+z_{s}\,\bar{z}_{v})+F(z_{s}\bar{z}_{s},\partial_{m}z_{s}\partial^{m}\bar{z}_{s},\ldots)\,,
\end{align}
where the index $m$ is raised and lowered with the Minkowski metric on the conformal boundary. As a comment, there can be finite covariant terms in $S_{ct}$ which, from the field theory point of view, reflect a potential dependence on the renormalisation scheme we can choose. We will further assume that we stick to a renormalisation scheme which respects the internal $U(1)$ symmetry of the bulk. Such scheme dependent contributions, when we replace the covariant fields by their UV expansion will be part of the function $F$ in \eqref{eq:Sct_fin}.

Including the finite terms, from the variation of the bulk action \eqref{eq:action_general} we get
\begin{align}
\langle \mathcal{O}_{Z}\rangle=(\Delta - \frac{d}{2})\,z_{v}+z_{s}\,\hat{F}(z_{s}\bar{z}_{s},\partial_{m}z_{s}\partial^{m}\bar{z}_{s},\ldots)\,,
\end{align}
where $\hat{F}$ can be written as a sum of derivatives of $F$. This shows that, in all cases we will subsequently consider, to leading order in the source, our one point functions will be given by e.g. $\langle \mathcal{O}_{Z}\rangle\approx (\Delta - \frac{d}{2})\,z_{v}$.

A second point that can be derived from the previous discussion is that explicit variations of the on-shell action \eqref{eq:action_total} with respect to $k_{j}$ in the backgrounds \eqref{eq:bh_sol} are going to come, to leading order in the source, from the bulk action \eqref{eq:action_general}. For this action, we have that the bulk contribution of the derivative of $w$ with respect to the wavelength $k_{i}$ is
\begin{align}
\partial_{k_{i}}w&=\int_{r_{h}}^{\infty}dr\,\sqrt{g}\,Q \rho^{2}g^{ij}k_{j}\,.
\end{align}
We can easily see that as long the scalar $z$ satisfies the unitarity bound $\Delta>\frac{d}{2}-1$, the above integral converges when $z_{s}=0$, and so the counterterm action does not contribute to this variation.

Assuming that $Q>0$, we see that the dominant branch of black holes will have $k_{j}=0$ in the class of theories described by \eqref{eq:action_general}. For this thermodynamically preferred branch of black holes, we can then easily argue that
\begin{align}\label{eq:w_second_der}
w^{ij}\equiv \left.\partial_{k_{i}}\partial_{k_{j}} w\right|_{k_{l}=0}&=\int_{r_{h}}^{\infty}dr\,\sqrt{g}\,Q \rho^{2}g^{ij}\,.
\end{align}
This discussion shows that if we perturb the backgrounds by a static source $\delta z_{s}$, the finite counterterms in $S_{ct}$ will contribute only to higher perturbative corrections of the free energy derivative $\partial_{k_{i}}\partial_{k_{j}}w\approx w^{ij}+\delta w^{ij}$, with $\delta w^{ij}$ proportional to the perturbative source $|\delta z_{s}|$.

\section{Linear response and the diffusive pole}\label{sec:diffusion}
In order to make the discussion of the hydrodynamic modes from the field theory side more transparent, we write our complex field in terms of real ones according to
\begin{align}\label{eq:z_def}
Z=X+i\,Y\,.
\end{align}
We now consider perturbations of $Y$ around the background \eqref{eq:bh_sol} with $k_{j}=0$. These are described by
\begin{align}\label{eq:eom_dy}
&\nabla_{\mu}\left( Q_{0}\,\nabla^{\mu}\delta Y\right)-Q_{u}\,(\partial \rho)^{2}\,\delta Y-V_{0}\,\delta Y=0\,,\nn
&Q_{0}=\left.Q\right|_{\mathcal{F}=\mathcal{F}_{0},Z=\rho},\quad Q_{u}=\left.\partial_{u}Q\right|_{\mathcal{F}=\mathcal{F}_{0},Z=\rho} ,\quad V_{0}=-\left.\partial_{u}W\right|_{\mathcal{F}=\mathcal{F}_{0},Z=\rho}\,.
\end{align}
In order to solve equation \eqref{eq:eom_dy}, we notice that for the background \eqref{eq:bh_sol} we have
\begin{align}
\nabla_{\mu}\left( Q_{0}\,\nabla^{\mu}\rho\right)-Q_{u}\,(\partial \rho)^{2}\,\rho-V_{0}\,\rho=0\,.
\end{align}
We now perform the change of fields
\begin{align}\label{eq:deltachi_def}
\delta Y=\rho\,\delta\chi\,,
\end{align}
which can be regarded as a small rotation in the complex plane where $Z$ takes values. The resulting equation of motion is
\begin{align}\label{eq:massless_eom}
\nabla_{\mu}\left( Q_{0}\rho^{2}\,\nabla^{\mu}\delta \chi\right)=0\,.
\end{align}
We observe that $\delta\chi=\delta c_{0}$ is a solution, and for the general solution at small frequency $\omega$ and wavelength $q_{j}$ we perform the expansion
\begin{align}\label{eq:hyd_exp}
\delta\chi(t,x^{j},r)&=e^{-i\,(\varepsilon_{\omega}\,\omega\,(t+S(r))-\varepsilon_{q}\,q_{j}x^{j})}\,\sum_{m,n=0}^{\infty}\varepsilon_{\omega}^{m}\varepsilon_{q}^{n}\,\delta\chi_{m,n}(r)\,,\notag\\
\delta\chi_{0,0}(r)&=\delta c_{0}\,.
\end{align}
Here $\varepsilon_{\omega}$ and $\varepsilon_{q}$ are small bookkeeping parameters. The ansatz and logic used in order to solve \eqref{eq:massless_eom} is indeed very similar to \cite{Donos:2017ihe}, the main difference being that we will include an external source and of course this time we will not find necessary to impose a dispersion relation $\omega(q^{i})$, as $\omega$ and $q^{i}$ will now be free parameters for our source. The function $S(r)$ is such that it vanishes sufficiently fast close to the conformal boundary and close to the horizon it behaves like $S(r)\to \frac{1}{4\pi T}\,\ln(r-r_{h})+\cdots$. With this choice, close to the horizon the time $t$ and the function $S$ combine to the infalling Eddington-Finkelstein coordinate $v=t+\frac{1}{4\pi T}\,\ln(r-r_{h})$. Additionally, note that infalling boundary conditions at the horizon require $\delta\chi_{m,n}$ to be analytic functions at $r=r_{h}$.

Plugging the expansion \eqref{eq:hyd_exp} in \eqref{eq:massless_eom} and keeping terms up to $\mathcal{O}(\varepsilon_{\omega},\varepsilon_{q}^{2})$, we find the equations
\begin{align}\label{eq:pert_eoms}
\partial_{r}\left(\sqrt{g}\,Q_{0}\rho^{2}U\partial_{r}\delta\chi_{0,1} \right)&=0\,,\nn
\partial_{r}\left(\sqrt{g}\,Q_{0}\rho^{2}U\partial_{r}\delta\chi_{1,0} \right)-i\omega\,\delta c_{0}\,\partial_{r}\left(\sqrt{g}\,Q_{0}\rho^{2}U\partial_{r}S \right)&=0\,,\nn
\partial_{r}\left(\sqrt{g}\,Q_{0}\rho^{2}U\partial_{r}\delta\chi_{0,2} \right)-\delta c_{0}\,\sqrt{g}\,Q_{0}\rho^{2}\,g^{ij}\,q_{i}q_{j}&=0\,.
\end{align}
Using the asymptotic expansion of equation \eqref{eq:ads_as} we find that close to infinity we must have
\begin{align}\label{eq:dchi_exp}
\delta\chi_{0,1}&=\delta\chi^{(0)}_{0,1}\,,\notag\\
\delta\chi_{1,0}&=\frac{i\,\omega}{(d-2\Delta)\,r^{d-2\Delta}}\,\frac{\sqrt{g_{h}}Q_{0}^{h}\rho_{h}^{2}}{\rho_{v}^{2}}\,\delta c_{0}+\cdots+\delta\chi_{1,0}^{(0)}+\cdots\notag\,,\\
\delta\chi_{0,2}&=-\frac{q_{i}q_{j}}{(d-2\Delta)\,r^{d-2\Delta}}\,\frac{w^{ij}}{\rho_{v}^{2}}\,\delta c_{0}+\cdots+\delta\chi_{0,2}^{(0)}+\cdots\,,
\end{align}
where the index $h$ denotes background quantities evaluated on the black hole horizon and $\delta\chi_{m,n}^{(0)}$ are constants of integration which we cannot fix. In order to clarify the importance of this expansion, we now go back to the perturbation $\delta Y$ which at leading order in $\varepsilon_{\omega}$ and $\varepsilon_{q}$ has the expansion
\begin{align}
\delta Y&=e^{-i\,(\omega\,t-q_{j}x^{j})}\left(\frac{ i\omega\,\sigma_{b}-w^{ij}q_{i}q_{j}}{(d-2\Delta)\,\rho_{v}}\,\frac{\delta c_{0}}{r^{d-\Delta}}+\cdots+\frac{\rho_{v}\,\left(\delta c_{0}+\cdots\right)}{r^{\Delta}}+\cdots\right)\,,\notag\\
\sigma_{b}&\equiv \sqrt{g_{h}}Q_{0}^{h}\rho_{h}^{2}\,,
\end{align}
where the dots include higher order corrections in $\varepsilon_{\omega}$ and $\varepsilon_{q}$. From this expansion we can find the retarded Green's function for the operator dual to $Y$ after identifying its source and one point function. As anticipated in subsection \ref{subsec:thermo}, we indeed see that the source is parametrically smaller than the $r^{-\Delta}$ term due to our hydrodynamic limit. Thus, the leading order retarded Green's function for the operator dual to $Y$ reads
\begin{align}\label{eq:G_undeformed}
G_{YY}(\omega, q^{i})=\frac{\rho_{v}^{2}\,(2\Delta-d)^{2}}{w^{ij}q_{i}q_{j}-i\omega\,\sigma_{b}}=\frac{\langle \mathcal{O}_{X}\rangle^{2}}{w^{ij}q_{i}q_{j}-i\omega\,\sigma_{b}}\,,
\end{align}
yielding a diffusive pole at $\omega=-i\,\sigma_{b}^{-1}\,w^{ij}q_{i}q_{j}$. Notice that $G(\omega=0,q^{i})$ diverges as $q^{i}\to0$, which is compatible with the existence of long range interactions from the field theory side point of view.

By Fourier transforming to spacetime coordinates, the result \eqref{eq:G_undeformed} for the retarded Green's function allows us to write an equation which determines the VEV $\delta \langle \mathcal{O}_{Y}\rangle(t,x^{i})$ of the operator $\mathcal{O}_{Y}$ in terms of the source $\delta z_{s}=i\,\zeta_{Y}(t,x^{i})$. After introducing the infinitesimally small angle $\delta\hat{c}_{g}(t,x^{i})$ through $\delta \langle \mathcal{O}_{Y}\rangle(t,x^{i})= \langle \mathcal{O}_{X}\rangle\,\delta\hat{c}_{g}(t,x^{i})$ we have
\begin{align}\label{eq:Josephson}
\left( \sigma_{b}\,\partial_{t} -w^{ij}\,\partial_{i}\partial_{j} \right)\delta\hat{c}_{g}=\langle \mathcal{O}_{X}\rangle\,\zeta_{Y}\,.
\end{align}
This equation provides an effective description for the long wavelength response of the emergent gapless mode due to the symmetry breaking in the bulk. In other words, we have derived a Josephson type of relation which is more familiar in superfluids. The main obvious difference from the phason mode of superfluids is that ours is a purely diffusive, non-propagating one. The new transport coefficient $\sigma_{b}$ that appears in \eqref{eq:Josephson} can be simply computed from the retarded Green's function after taking into account the right order of limits
\begin{align}
\frac{\langle \mathcal{O}_{X}\rangle^{2}}{\sigma_{b}}=\lim_{\omega \to 0}\lim_{q \to 0}\left(-i\,\omega\, G_{YY}(\omega,q^{i})\right)\,.
\end{align}
The last expression relates the transport coefficient $\sigma_{b}$ to the rate at which the system absorbs energy at long wavelengths and small frequencies.

\section{Explicit breaking and the gap}\label{sec:gap}
In this section we will consider once again the thermal state described by the general bulk geometry \eqref{eq:bh_sol}, this time right below the critical temperature $T_{c}$. In that regime $\langle \mathcal{O}_{X}\rangle\neq 0$ and $\rho$ is non-trivial in the bulk with asymptotics given by \eqref{eq:ads_as}. In the case where $\langle \mathcal{O}_{Z}\rangle$ is relevant, we can introduce an explicit source $\delta \rho_{s}$ in order break the bulk $U(1)$ symmetry in a controlled, perturbative fashion. With this new scale in our theory, it is useful to define a new bookkeeping dimensionless parameter $\delta\rho_{s}\approx \mathcal{O}(\varepsilon_{s})$ while once again we take $\omega \approx \mathcal{O}(\varepsilon_{\omega})$ and $q_{i}\approx \mathcal{O}(\varepsilon_{q})$. We will examine the retarded Green's function of $Y$ in the regimes $\varepsilon_{s}\gg\varepsilon_{\omega},\varepsilon_{q}$ and $\varepsilon_{s}\approx \varepsilon_{\omega} \approx \varepsilon_{q}^{2}$.

The perturbative deformation will modify the asymptotic expansion \eqref{eq:ads_as} to
\begin{align}\label{eq:explicit_expansion}
\hat{\rho}=\rho +\delta\rho=\frac{\delta\rho_{s}}{r^{d-\Delta}}+\cdots+\frac{\rho_{v} +\delta\rho_{v}}{r^{\Delta}}+\cdots\,,
\end{align}
where $\delta\rho_{v}$ is the perturbation of the VEV due to the addition of the perturbative source. Once again, we wish to study perturbations of $Y$, the imaginary part of $Z$, through the fluctuations of the field $\delta\chi$ which still satisfy the bulk equation \eqref{eq:massless_eom}. It is easy to see that $\delta\chi=\delta c_{0}$ being a constant provides a solution with asymptotic expansion
\begin{align}\label{eq:constant_mode}
\delta Y=\frac{\delta\rho_{s}}{r^{d-\Delta}}\,\delta c_{0}+\cdots+\frac{\rho_{v}+\delta\rho_{v}}{r^{\Delta}}\,\delta c_{0}+\cdots\,.
\end{align}
Recalling the discussion in subsection \ref{subsec:thermo}, this suggests that in the perturbatively deformed theory we have the thermodynamic susceptibility
\begin{align}
\chi_{YY}\equiv G_{YY}(\omega=0,q^{i}=0)=(2\Delta -d)\frac{\rho_{v}}{\delta\rho_{s}}=\frac{\langle \mathcal{O}_{X}\rangle}{\delta\rho_{s}}\,.
\end{align}

We now find a mode close to the static, homogeneous solution \eqref{eq:constant_mode}, which has frequency $\omega$ and wavenumber $q^{2}$ at the scale set by $\delta\rho_{s}$. In order to do this, we perform an expansion in $\varepsilon_{\omega}$, $\varepsilon_{q}$ and $\varepsilon_{s}$, similar to \eqref{eq:hyd_exp}
\begin{align}
\delta\chi(t,x^{j},r)&=e^{-i\,(\varepsilon_{\omega}\,\omega\,(t+S(r))-\varepsilon_{q}\,q_{j}x^{j})}\,\sum_{m,n,l=0}^{\infty}\varepsilon_{\omega}^{m}\varepsilon_{q}^{n}\varepsilon_{s}^{l}\,\delta\chi_{m,n,l}(r)\,,\notag\\
\delta\chi_{0,0,0}(r)&=\delta c_{0}\,.
\end{align}
At this point we stress that in order to solve the equation of motion \eqref{eq:massless_eom} perturbatively in beyond next to leading order in $\varepsilon_{s}$, we would need to know the full perurbative expansion of the background. Luckily, in order to extract non-trivial information, we will only need to work at leading order in $\varepsilon_{s}$. We can indeed see that, at leading order in $\delta\rho_{s}$, we arrive again at a set of equations identical to \eqref{eq:pert_eoms} for $\delta\chi_{0,1,0}$, $\delta\chi_{1,0,0}$ and $\delta\chi_{0,2,0}$ with solution
\begin{align}\label{eq:dchi_exp_def}
\delta\chi_{0,1,0}&=\delta\chi^{(0)}_{0,1,0}\,,\notag\\
\delta\chi_{1,0,0}&=\frac{i\,\omega}{(d-2\Delta)\,r^{d-2\Delta}}\,\frac{\sqrt{g_{h}}Q_{0}^{h}\rho_{h}^{2}}{\rho_{v}^{2}}\,\delta c_{0}+\cdots+\delta\chi_{1,0,0}^{(0)}+\cdots\notag\,,\\
\delta\chi_{0,2,0}&=-\frac{q_{i}q_{j}}{(d-2\Delta)\,r^{d-2\Delta}}\,\frac{w^{ij}}{\rho_{v}^{2}}\,\delta c_{0}+\cdots+\delta\chi_{0,2,0}^{(0)}+\cdots\,.
\end{align}
Thus, at order $\varepsilon_{s}\approx \varepsilon_{\omega} \approx \varepsilon_{q}^{2}$ that we are interested in, we have
\begin{align}
\delta Y&=e^{-i\,\left(\omega\,t -q_{j}x^{j}\right)}\left( \left(\delta\rho_{s} + \frac{ i\omega\,\sigma_{b}-w^{ij}q_{i}q_{j}}{(d-2\Delta)\,\rho_{v}}\right)\,\frac{\delta c_{0}}{r^{d-\Delta}}+\cdots+\frac{\rho_{v}\,\left(\delta c_{0}+\cdots\right)}{r^{\Delta}}+\cdots\right)\,,
\end{align}
where the dots include higher order corrections in $\varepsilon_{s}$. This expansion constitutes yet another example of the discussion in subsection \ref{subsec:thermo} and so we can read off the retarded Green's function to be
\begin{align}\label{eq:G_deformed}
G_{YY}(\omega, q^{i})=\frac{\langle \mathcal{O}_{X}\rangle^{2}}{\langle \mathcal{O}_{X}\rangle\,\delta\rho_{s}+w^{ij}q_{i}q_{j}-i\omega\,\sigma_{b}}=\frac{\langle \mathcal{O}_{X}\rangle^{2}}{\langle \mathcal{O}_{X}\rangle^{2}\,\chi_{YY}^{-1}+w^{ij}q_{i}q_{j}-i\omega\,\sigma_{b}}\,.
\end{align}
yielding a quasinormal mode with a gap $\delta\omega_{g}$ given by\footnote{We believe that at leading order in their ``pinning'' parameter, this new result should match the formula for $\Omega$ given as a bulk integral in the notation of \cite{Amoretti:2018tzw}.}
\begin{align}\label{eq:gap}
\delta\omega_{g}=\frac{\langle \mathcal{O}_{X}\rangle}{\sigma_{b}}\,\delta\rho_{s}=\frac{\langle \mathcal{O}_{X}\rangle^{2}}{\sigma_{b}\,\chi_{YY}}\,.
\end{align}
After this definition, the Josephson relation \eqref{eq:Josephson} for the angular variable $\delta\hat{c}_{g}$ is modified to
\begin{align}\label{eq:Josephson_gapped}
\left[ \sigma_{b}\,(\partial_{t}+\delta\omega_{g}) -w^{ij}\,\partial_{i}\partial_{j} \right]\delta\hat{c}_{g}=\langle \mathcal{O}_{X}\rangle\,\zeta_{Y}\,,
\end{align}
demonstrating the fact that $\delta\omega_{g}$ can be thought of as a relaxation time for the perturbatively gapped mode.

In the next section we will provide a numerical check of our claim for a specific model. Here we will also provide a check of equation \eqref{eq:G_deformed} at small frequencies and wavenumbers, much smaller than the scale set by $\delta\rho_{s}$. We are once again performing an expansion similar to \eqref{eq:hyd_exp} which leads to \eqref{eq:pert_eoms} but now with $\rho$ being replaced by $\hat{\rho}$. At leading order in $\delta\rho_{s}$ we have
\begin{align}\label{eq:dchi_exp_def_hydro}
\delta\chi_{0,1}&=0\,,\notag\\
\delta\chi_{1,0}&=-\frac{i\,\omega}{(d-2\Delta)\,r^{2\Delta-d}}\,\frac{\sigma_{b}}{\delta\rho_{s}^{2}}\,\delta c_{0}+\cdots\notag\,,\\
\delta\chi_{0,2}&=\frac{q_{i}q_{j}}{(d-2\Delta)\,r^{2\Delta-d}}\,\frac{w^{ij}}{\delta\rho_{s}^{2}}\,\delta c_{0}+\cdots\,.
\end{align}
The corresponding asymptotic expansion for the perturbation $\delta Y$ reads
\begin{align}\label{eq:Y_asympt_exp_def}
\delta Y&=e^{-i\,(\omega\,t-q_{j}x^{j})}\left[\frac{\delta\rho_{s}}{r^{d-\Delta}}+\cdots+\frac{1}{(d-2\Delta)\,r^{\Delta}}\left((d-2\Delta) \rho_{v}-i\omega\frac{\sigma_{b}}{\delta\rho_{s}}+\frac{w^{ij}q_{i}q_{j}}{\delta\rho_{s}}\right)+\cdots\right]\,\delta c_{0}\,,
\end{align}
where the dots include higher order corrections in $\varepsilon_{s}$ as well as in $\varepsilon_{\omega}$ and $\varepsilon_{q}$. The expansion \eqref{eq:Y_asympt_exp_def} yields the expanded Green's function
\begin{align}
G_{YY}(\omega, q^{i})\approx \chi_{YY}+ \frac{\langle \mathcal{O}_{X}\rangle}{\delta\rho_{s}} \frac{i \omega - w^{ij}q_{i}q_{j} / \sigma_b}{\delta \omega_g} +\cdots\,,
\end{align}
which is simply an expansion of \eqref{eq:G_deformed} for $\varepsilon_{\omega}\approx\varepsilon_{q^{2}}\ll \varepsilon_{s}$.

\section{Numerical checks}\label{sec:numerics}

The aim of this section is to verify numerically the analytic results of sections \ref{sec:diffusion} and \ref{sec:gap} in a simple theory in $d=3$ boundary dimensions. The action that we will consider is given by
\begin{align}\label{eq:action}
S=\int d^4 x \sqrt{-g}\,\left[R-V-\frac{1}{2}\,\partial^\mu\psi \, \partial_\mu\psi -\frac{1}{2}\,\partial^\mu\rho \, \partial_\mu\rho -\frac{\rho^2}{2}(\partial \chi)^2\right]\,,
\end{align}
where we have chosen
\begin{align}
&V=-6+\frac{m_\psi^2}{2}\psi^2+\frac{c^2}{2}\psi^4+\frac{m_\rho^2}{2}\phi^2+ q\,\rho^2\psi^2\,,\quad m_\psi^2=-2\,,\quad m^2_\phi=-2\,.
\end{align}
In comparison to \eqref{eq:action_general} in section \ref{sec:setup},  $Z=\rho\, e^{i \chi}, Q(Z\bar{Z}\,,\mathcal{F})=1$ and $W$ corresponds to the first three terms in our action. Note that, for $\rho=0$, $\partial_\psi V=0$ both at $\psi=0$ and at $\psi=1/c$. 

The variation of the action \eqref{eq:action} gives rise to the following field equations of motion
\begin{align}
\label{eq:eom}
&R_{\mu\nu}-\frac{1}{2}g_{\mu\nu} V-\frac{1}{2}\partial_\mu\rho\partial_\nu\rho-\frac{1}{2}\partial_\mu\psi\partial_\nu\psi-\frac{\rho^2}{2}\partial_\mu\chi\partial_\nu\chi=0\,,\nonumber\\
& \frac{1}{\sqrt{-g}}\partial_\mu\left(\sqrt{-g}\,\partial^\mu\psi\right)-\partial_{\psi}V=0\,,\nonumber\\
&\frac{1}{\sqrt{-g}}\partial_\mu\left(\phi^2\sqrt{-g}\,\partial^\mu\chi_i\right)=0\,,\nonumber\\
& \frac{1}{\sqrt{-g}}\partial_\mu\left(\sqrt{-g}\,\partial^\mu\rho\right)-\partial_{\rho}V-\rho\,(\partial \chi)^{2}=0\,.
\end{align}
The above equations admit a unit-radius $AdS_4$ vacuum solution with $\rho =\psi= \chi=0$, which is dual to a $d = 3$ CFT. Placing the CFT at finite temperature corresponds to considering the Schwarzschild black hole. 
%\begin{align}\label{eq:Schw}
%&U= (r+R)^{2}-\left(R^{2}+\frac{\mu^{2}}{4} \right)\frac{R}{r+R},\quad V_{1}=\ln (r+R),\notag\\
%&R=\frac{4}{3}\pi \,T\,, \quad \phi=0\,,\quad \psi=0\,.
%\end{align} 
Here we choose to deform our boundary theory by a relevant operator $\mathcal{O}_\psi$ with scaling dimension $\delta_\psi= 2$. The corresponding back-reacted solution will then be given by black holes with a non-trivial profile for the scalar field $\psi$. As the temperature goes to zero, $T\to 0$, these configurations will approach a flow between the unit-radius $AdS_4$ in the UV with $\psi=0$, and an IR $AdS_4$ with radius $L_{IR}^2=12 c^2/(1+12 c^2)$ supported by $\psi=1/c$. 

To see this explicitly, we consider the following ansatz  for the normal phase of our system 
\begin{align}\label{eq:ansatz_np}
ds^{2}&=-U(r)\,dt^{2}+\frac{1}{U(r)}\,dr^{2}+e^{2V(r)}\,dx_i^2\,,\nn
\psi&=\psi(r)\,,\quad \rho=0\,,\quad \chi=0\,,
\end{align}
where $i=1,2$. Plugging this in the equations of motion \refeq{eq:eom} we obtain one first order and two second order ODEs, which we solve numerically using a shooting method, subject to the following boundary conditions at the horizon
\begin{align}\label{eq:IRexp_np}
U\left(r\right)&=4\pi\,T\,(r-r_h)+\cdots\,,\qquad V=V_h+\cdots\,,\nn
\psi=&\psi_h+\cdots\,,
\end{align}
and the following boundary behaviour
\begin{align}\label{eq:UVexp_np}
U&\to r^2+\cdots+\frac{W}{r}+\cdots,\qquad V\to \log (r)+\cdots\,,\nn
\psi&\to \frac{\psi_s}{r}\,+\frac{\psi_{v}}{r^{2}}+\cdots\,.
\end{align}
Note that, in the above expansion, $T$ is the black hole temperature and $\psi_s$ is the deformation parameter.

Given the above expansions, and through a simple counting argument, we expect to find a two-parameter family of black holes labelled by $(T,\psi_s)$.

\begin{figure}[h!]
\centering
\includegraphics[width=0.48\linewidth]{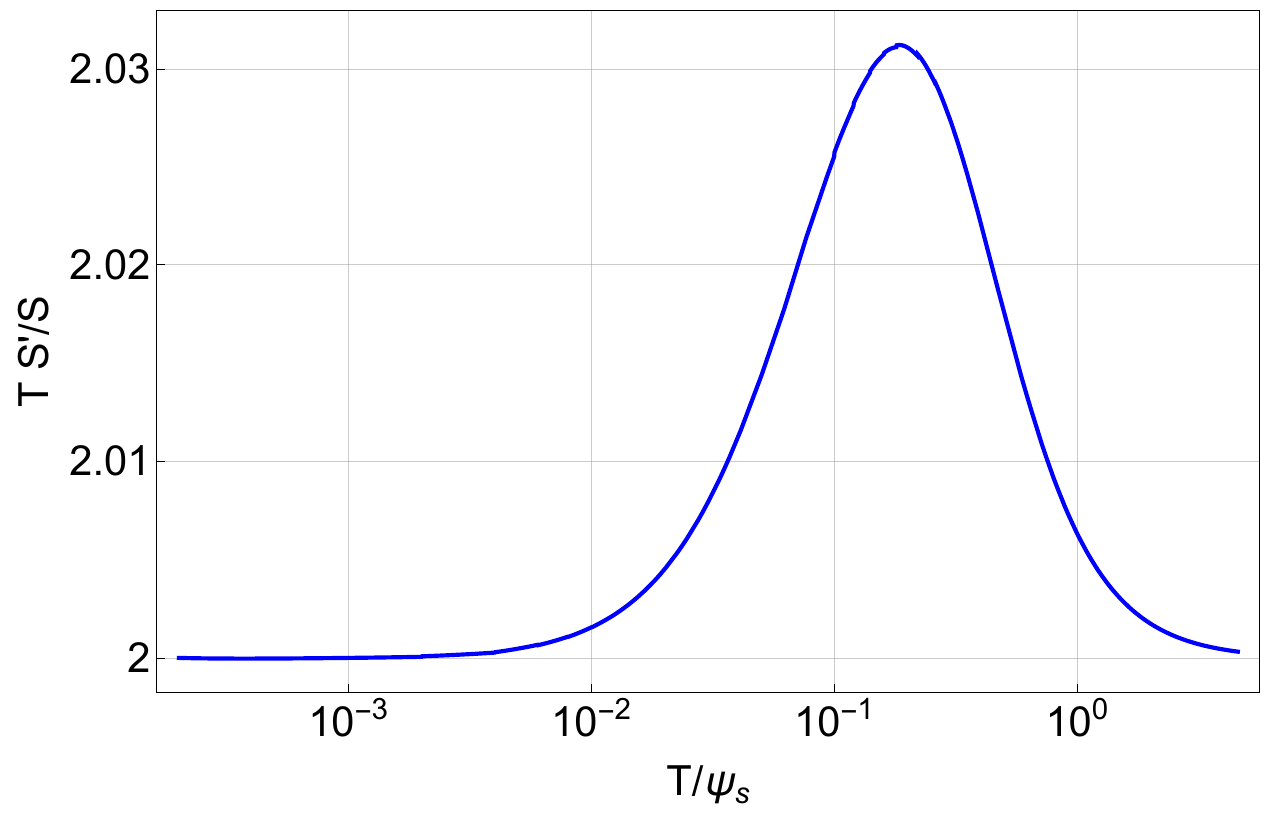}\quad\includegraphics[width=0.48\linewidth]{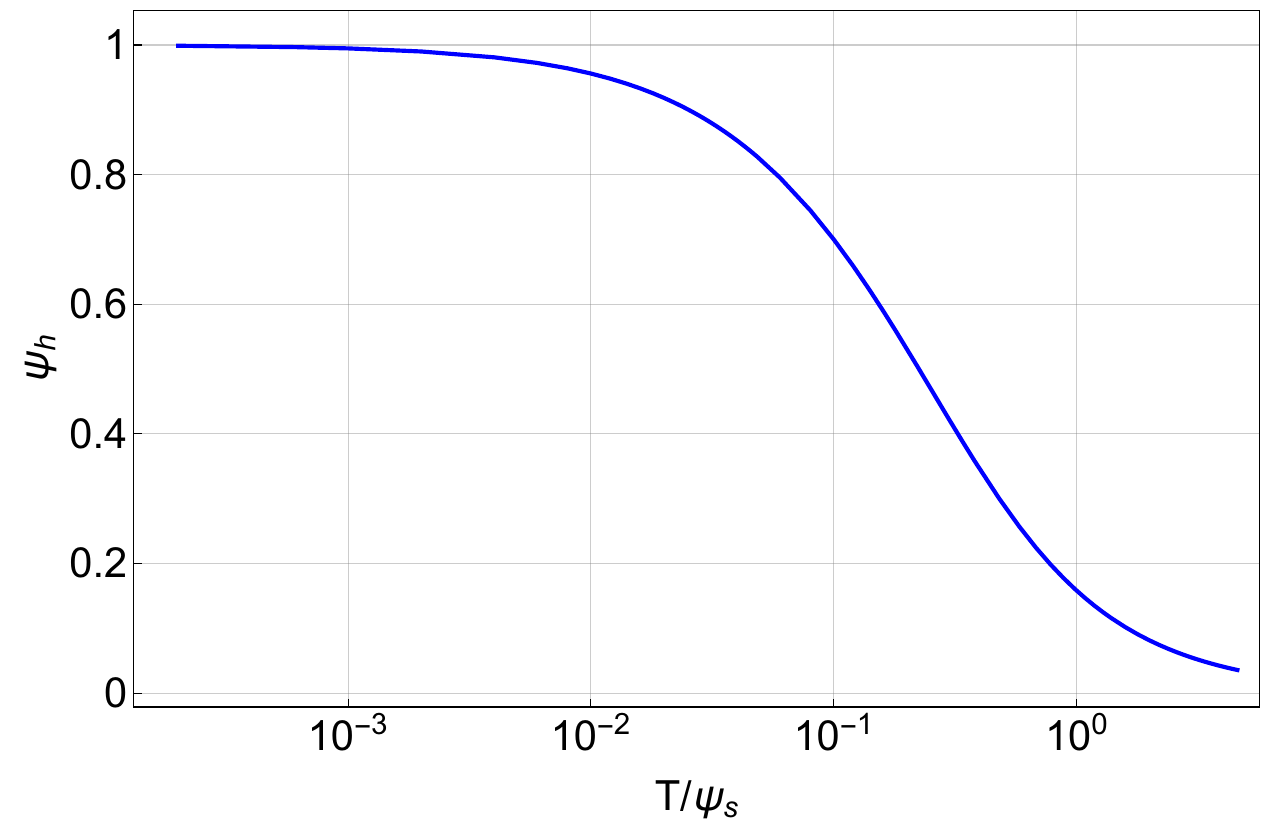}
\caption{The left panel shows the logarithmic derivative of the entropy of the system with respect to the temperature, while the right panel shows the horizon constant $\psi_h$ as a function of the temperature. Here $\psi_s=1/2$ and $c=1,q=-1$.}
\label{fig:NormalPhase}
\end{figure}

In the left panel of Figure \ref{fig:NormalPhase}, we plot the logarithmic derivative of the entropy of the system with respect to the temperature, $T S'(T)/S(T)$ for $\psi_s=1$ and $c=1,q=-1$.  We clearly see that both at very high and very low temperatures the entropy scales like $T^2$ which is compatible with having $AdS_4$ on both sides of the RG flow. Furthermore, in the right panel of \ref{fig:NormalPhase} we plot $\psi_h$ against the temperature and we see that, in the deep IR, $\psi\to 1$ as expected when $c=1$.

\begin{figure}[h!]
\centering
\includegraphics[width=0.48\linewidth]{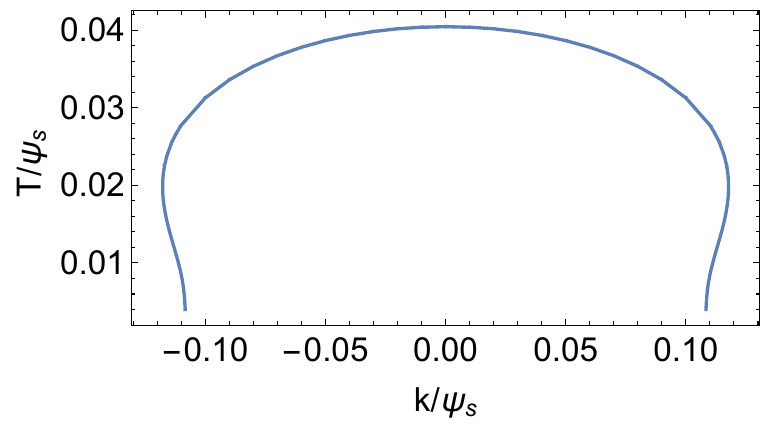}
\caption{Plot of the critical as a function of the wavenumber $k$.}
\label{fig:Tc}
\end{figure}

On top of these thermal states, we will consider instabilities associated to the scalar field $\rho$ which plays the role of the real part of the scalar $Z$ in section \ref{sec:setup}. To ensure that such instabilities exist in our model we need to make sure that the scalar field $\rho$ violates the BF bound associated with the $AdS_4$ in the IR, i.e.
\begin{equation}
L^2_{IR} \,{m^{IR}_\rho}^2=\frac{12 c^2}{1+12 c^2}(m_\rho^2+2\frac{q}{c^2})< -\frac{9}{4}\,,
\end{equation}
but is nevertheless stable in the UV: $m_\rho^2\ge-\frac{9}{4}\,$. For example, this is the case for $c=1,q=-1$ and so we expect an instability to occur for this choice of parameters. Thus, for temperatures below a critical one, we expect a new branch of black holes to emerge characterised by a non-trivial condensate for $\rho$. To determine the critical temperature at which these instabilities set in we need to study the associated zero modes. In particular, we consider a linearised perturbation around the background \eqref{eq:ansatz_np} of the form
\begin{align}
\rho(r,x_1)&=\delta\rho(r)\, \cos{(k\, x_1)}\,,
\end{align}    
where $k$ is the wavenumber that characterises the broken, generically spatially modulated phase. Note that in this paper we are mainly interested in the translationally invariant case with $k=0$; here we present the non-zero $k$ branches just for completeness. Plugging the above perturbation in the equations of motion, we obtain the following second order ODE 
\begin{align}\label{eq:rho_zeromode}
\delta\rho''+\frac{\rho}{U}(2+2 \psi^2-k^2 e^{-2V})+\frac{\rho'}{4 V'\, U}\left(12+2\psi^2-\psi^4+U(4 V'^2+\psi'^2\right)=0\,.
\end{align}
To solve the latter, we impose the following boundary conditions at the black hole horizon
\begin{align} 
\delta \rho&=\delta\rho_h+\cdots\,,
\end{align}  
and asymptotically
\begin{align}
\delta \rho&=0+\frac{\delta\rho_v}{r^2}+\cdots\,,
\end{align}  
where we have already set the source for $\delta \rho$ to $0$, so that the emergence of the new phase is spontaneous. Overall, the boundary conditions are determined by 2 constants $\delta\rho_h=1\,,\delta\rho_v$ one of which can be set to $1$ because of the linearity of equation \eqref{eq:rho_zeromode}. Therefore, for a fixed value of $k$ there will be at most a discrete set of temperatures for which equation \eqref{eq:rho_zeromode} will admit a solution without a boundary source. Consequently, performing the numerical integration for different values of $k$ will fix the highest value of the temperature for which one can find non-trivial solutions for $\delta\rho$. In Figure \ref{fig:Tc} we plot this critical temperature as a function of $k$. We indeed see that the dominant branch will correspond to $k=0$ as it is the first black hole that will appear as we lower the temperature. This certainly agrees with the analysis of the non-linear backgrounds that we discussed in subsection \ref{subsec:thermo}. We also note that the critical temperature displays the usual Bell curve behaviour \cite{Nakamura:2009tf,Donos:2011bh,Donos:2011qt} only for small values of $k$. For larger values of $k$ we see that it deviates from that form due to the fact that the near horizon geometry describing the zero temperature limit of the normal phase black holes is $AdS_{4}$ rather than $AdS_{2}\times \mathbb{R}^{2}$.

The next step is to construct the back-reacted solutions corresponding to the broken phase for $k=0$. This is achieved using the following ansatz
\begin{align} \label{eq:DC_ansatz}
ds^{2}&=-U(r)\,dt^{2}+\frac{1}{U(r)}\,dr^{2}+e^{2V(r)}\,dx_i^2\,,\nn
\rho&=\rho(r)\,,\quad \psi=\psi(r)\,,\quad \chi=0\,,
\end{align}
where $i=1,2$. When plugging this ansatz in the equations of motion we obtain a set of three second-order equations and one first-order equation. Thus, a solution is specified in terms of 7 constants of integration. To ensure that we have a regular Killing horizon we assume that we have the following expansions near the horizon
\begin{align}\label{eq:IRexp_bh}
U\left(r\right)&=4\pi\,T\,(r-r_h)+\cdots\,,\qquad V=V_h+\cdots\,,\nn
\rho&=\rho_h+\cdots\,,\qquad \psi=\psi_h+\cdots\,.
\end{align}
The expansion \eqref{eq:IRexp_bh} is then specified in terms of 4 constants, in addition to the temperature $T$. Asymptotically, we have the following expansion 
\begin{align}\label{asymptsol}
U&\to r^2+\cdots+\frac{W}{r}+\cdots,\qquad V\to \log(r)+\cdots\,,\nn
\rho&\to 0+\frac{\rho_v}{r^{2}}+\cdots, \qquad \psi\to  \frac{\psi_s}{r}\,+\frac{\psi_{v}}{r^{2}}+\cdots\,.
\end{align}
Note that, just like in the discussion for the zero modes, we have fixed the source for $\rho$ to be zero $\rho_s=0$, so that the condensation is spontaneous. The asymptotics are parametrised by 3 constants in addition to $\psi_s$. Overall, in the IR and UV expansions we have a total of 7 constants as well as $T$ and $\psi_s$, which matches the 7 constants of integration. We proceed to solve this boundary condition problem numerically using double-sided shooting for $(T, \psi_s)=(0.02,1)$ and  $c=1,q=-1$. 

\subsection{Diffusive mode and gap}

Having constructed the back-reacted black holes, we now turn our attention to studying the two-point function $G_{Y\,Y}(\omega,q)$ and the gap $\delta \omega_g$ associated with these thermal states; the real field $Y$ was defined in \eqref{eq:z_def} as $Y=Im[Z]$.
%In this section we compute the spatially resolved quasinormal modes for the backgrounds constructed in the previous subsection.  
To do this, we consider the following linearised perturbation
\begin{equation}
\delta\chi(t,r,x_1)=e^{-i \omega v(t,r)+i q x_1}\delta\chi(r)\,,
\end{equation}
where $v$ is the infalling Eddington-Finkelstein coordinate defined as
\begin{align}
v(t,r)=t+\int_{\infty}^{r}\frac{dy}{U(y)}\,,
\end{align}
thus fixing the function $S(r)$ used in sections \ref{sec:diffusion} and \ref{sec:gap}. Note that we picked the momentum $q$ to point in the direction $x_1$ without loose of generality given that the background is isotropic for $k=0$. Plugging this ansatz in the equations of motion, we obtain one second order ODE. We now turn to the boundary conditions for these functions. In the IR we impose infalling boundary conditions at the horizon at $r=r_h$
\begin{align}
&\delta\chi=c_{IR}+\cdots\,,
\end{align}
where $c_{IR}$ is a constant. Thus, for fixed value of $q$ and $\omega$, we see that the expansion is fixed in terms of 1 constant. The UV expansion for $\chi$ changes depending on whether we have $\rho_s=0$ (needed for the computation of $G_{Y\,Y}(\omega,q)$) or $\rho_s\ne0$ (needed for the computation of $\delta \omega_g$), so we will discuss them separately. In both cases we proceed to solve this equation numerically subject to the appropriate boundary conditions using a shooting method. 

\subsubsection{Computing the two-point function}
\begin{figure}[h!]
\centering
\includegraphics[width=0.48\linewidth]{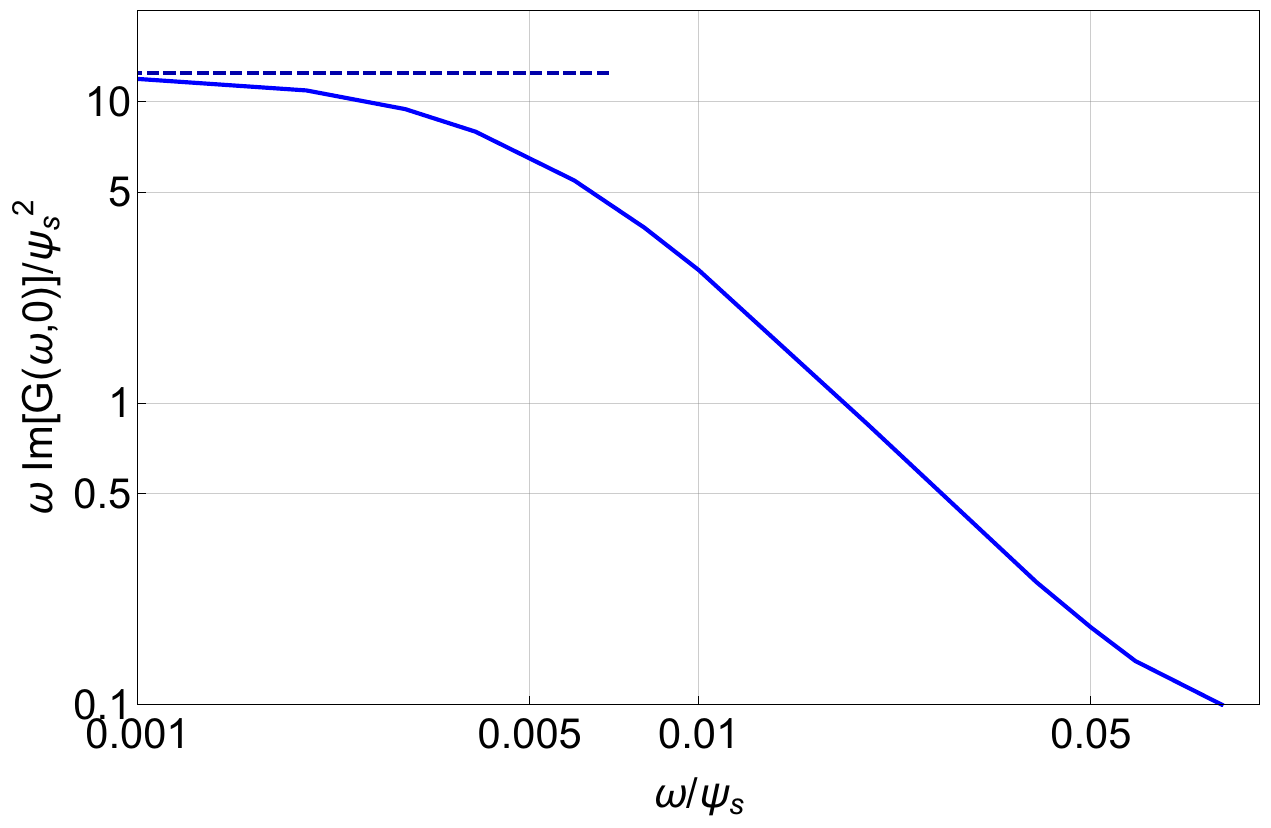}\quad \includegraphics[width=0.48\linewidth]{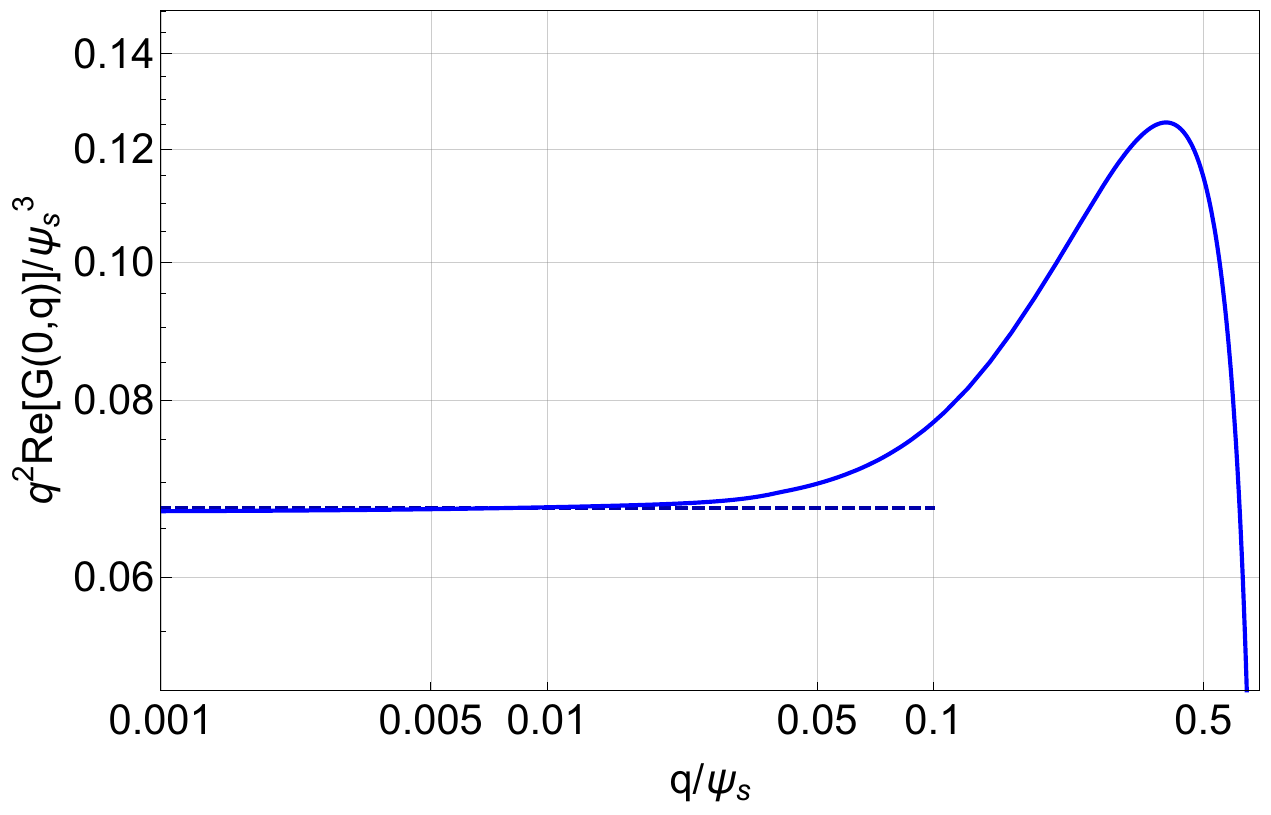}
\caption{We compare the analytic (dashed) and the numerical (solid) results for the two point functions $G(\omega,0)$ and $G(0,q)$ as a function of the frequency and the wavevector.}
\label{fig:GF}
\end{figure}

\begin{figure}[h!]
\centering
\includegraphics[width=0.48\linewidth]{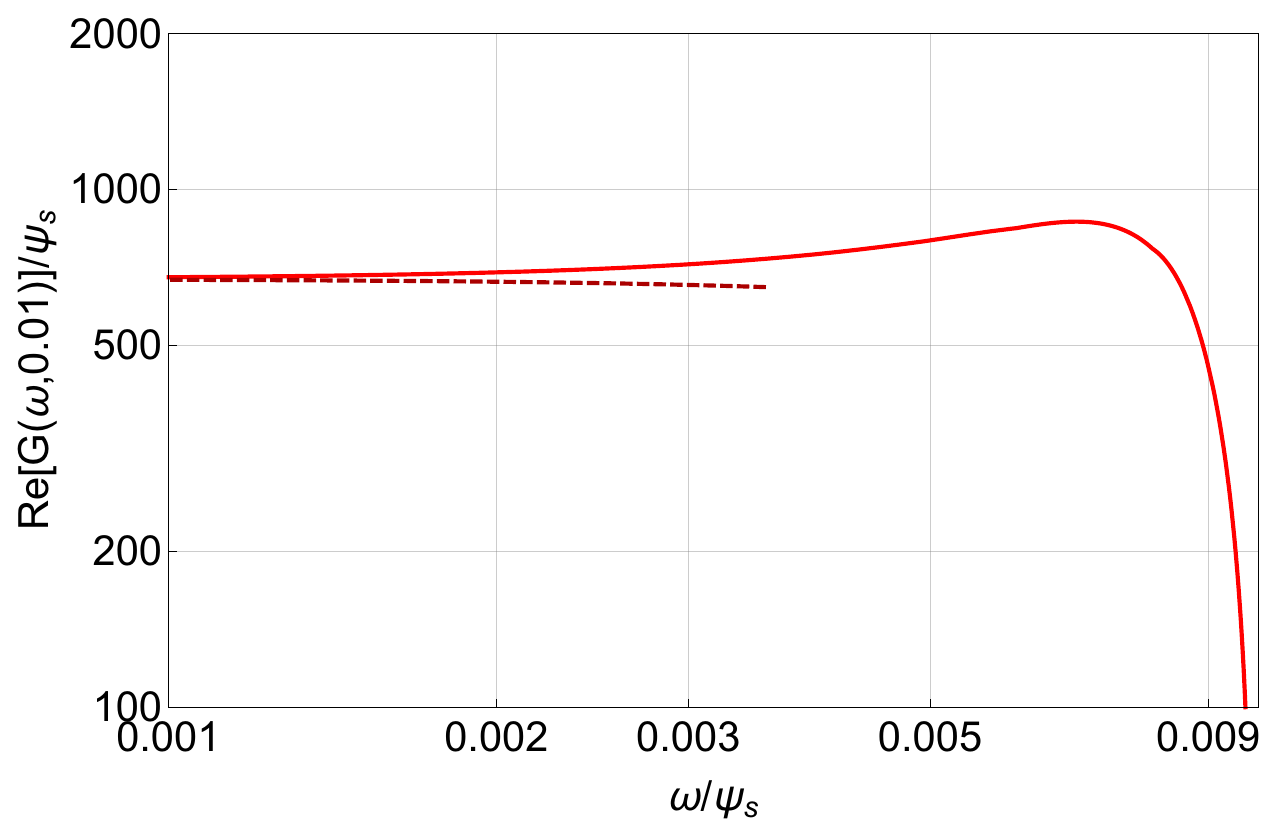}\quad \includegraphics[width=0.48\linewidth]{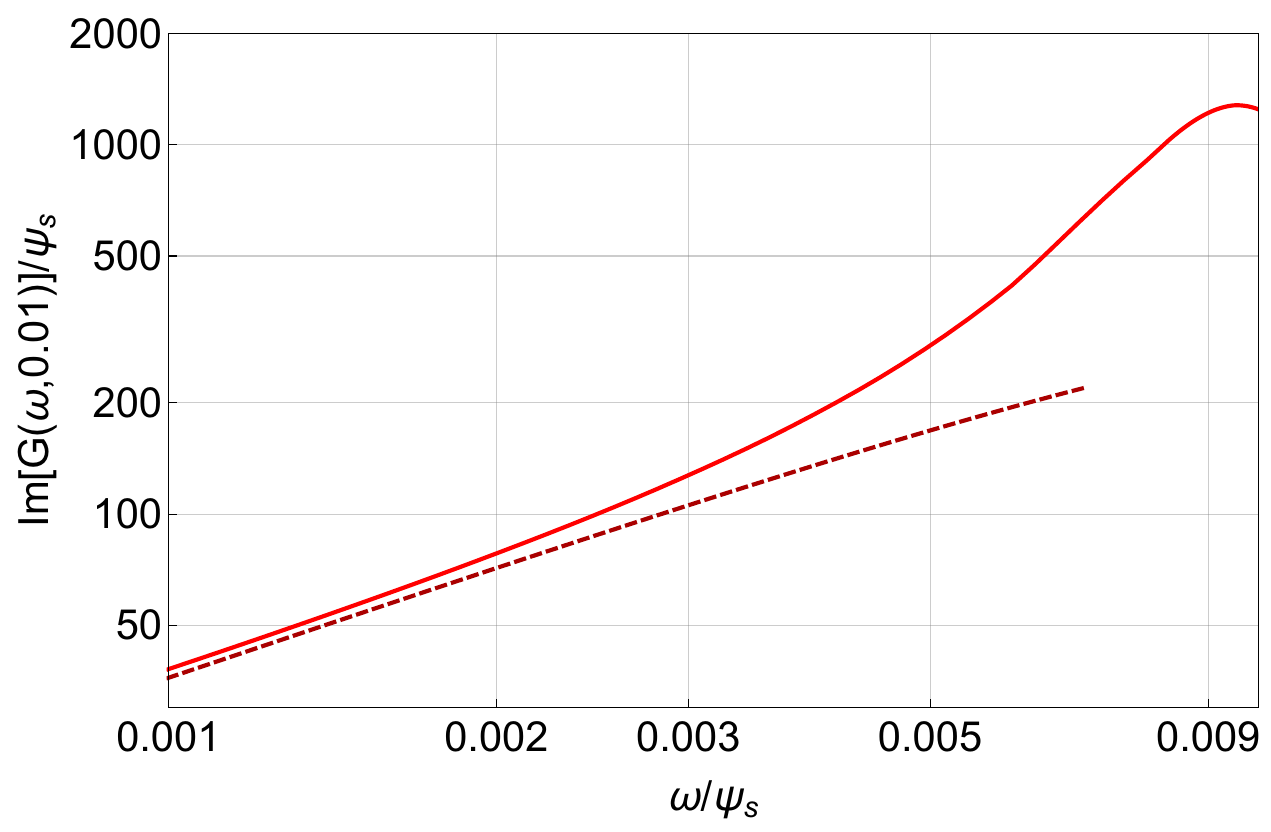}\\
\caption{We compare the analytic (dashed) and the numerical (solid) result for the two point function $G(\omega,0.01)$ for real frequencies.}
\label{fig:GF1}
\end{figure}

\begin{figure}[h!]
\centering
\includegraphics[width=0.48\linewidth]{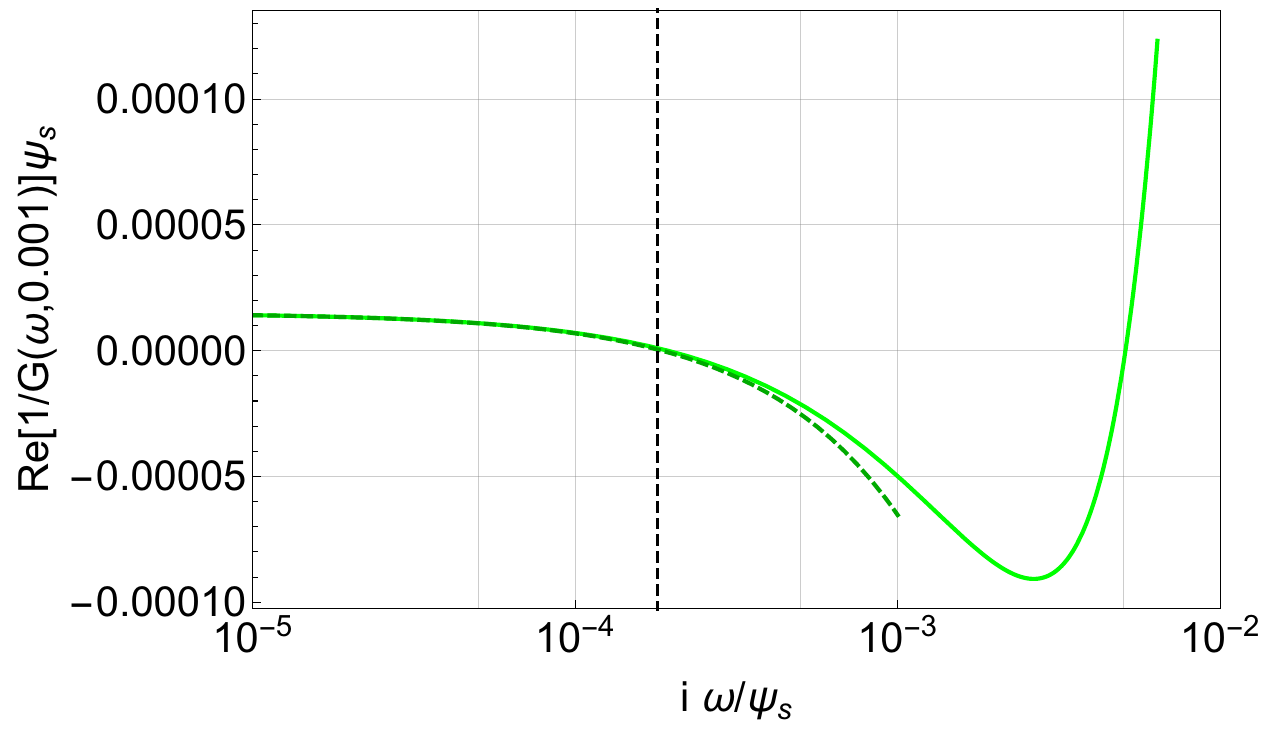}
\caption{We compare the analytic (dashed) and the numerical (solid) result for the two point function $G(\omega,0.001)$ for purely imaginary frequencies.}
\label{fig:GF2}
\end{figure}

When $\rho_s=0$ in the background, the UV expansion for the perturbation takes the form
\begin{align}
\label{eq:pertUV_2pf}
&\delta\chi=r \,\delta\chi_s+\delta\chi_v+\cdots\,.
\end{align}
In this case the parameter counting goes as follows. For fixed $\omega,q$, the UV and IR expansions are determined by $\delta\chi^{(v)}$ and $c_{IR}$, where we have already used linearity to set one of the parameters to unit $\delta\chi^{(s)}=1$. This matches the 2 integration constants of the problem.
After numerical integration, we proceed to calculate the two-point function
\begin{align}
G_{Y\, Y}\equiv G=\frac{\rho_v \,\delta\chi_v}{\rho_v\,\delta\chi_s}\,.
\end{align}
In Figure \ref{fig:GF} we plot $\omega \, \text{Im}[G(\omega,0)]/\psi_{s}^{2}$ and $q^2 \, \text{Re}[G(0,q)]/\psi_{s}^{3}$ as functions of the (dimensionless) frequency $\omega/ \psi_s$ and the (dimensionless) wavevector $q/\psi_s$ respectively. The dashed lines in the two plots correspond to the analytic result \eqref{eq:G_undeformed}, given by $\frac{\rho_v^2}{\sigma_b}=\frac{\rho_v^2}{e^{2V_h} \rho_h^2}=12.47$ and  $\frac{\rho_v^2}{w^{1\,1}}=0.067$, showing good agreement. Note that $w^{1\,1}$ was computed by performing explicitly the bulk integral
\begin{equation}
w^{1\,1}=\int_{r_h}^\infty dr \rho^2\,,
\end{equation}
defined in equation \eqref{eq:w_second_der}. For completeness, here we explicitly give all the constants that appear in the analytic expressions: $w^{1\,1}= 0.807\,, \rho_v=0.233\,, \sigma_b=0.0043$. In Figure \ref{fig:GF1} we plot the real and imaginary part of the $G(\omega, 0.01)$ for $\omega$ taking values in the reals; once again we see good agreement with the analytic results captured by the dashed lines. Finally, in Figure \ref{fig:GF2}, we plot the real part of $G(\omega,0.001)$ as a function of imaginary frequency $\omega$ for $q=1/1000$ and we see explicitly the appearance of a pole at $\omega=0.00018$.


\subsubsection{Computing the gap}
The computation of the gap boils down to computing quasinormal modes with $q=0$ in the presence of a small explicit breaking in the background, $\delta\rho_s\ne0$, also known as pinning. Having a non-vanishing source for the scalar field $\rho$ leads to the following UV expansion for the perturbation
\begin{align}
\label{eq:pertUV_gap}
&\delta\chi_1=\delta\chi_s+\frac{\delta\chi_v}{r}+\cdots\,,
\end{align}
and we need to remove the source by setting $\delta\chi^{(s)}=0$ from the UV expansion as prescribed for the computation of quasinormal modes. Thus, we see that the UV expansion is fixed in terms of 1 constant. Overall, in this boundary condition problem we have 2 undetermined constants $\omega,\delta\chi^{(v)}$ together with $c_{IR}=1$ due to the linearity of the equation. This matches precisely the 2 integration constants of the problem.

\begin{figure}[h!]
\centering
\includegraphics[width=0.68\linewidth]{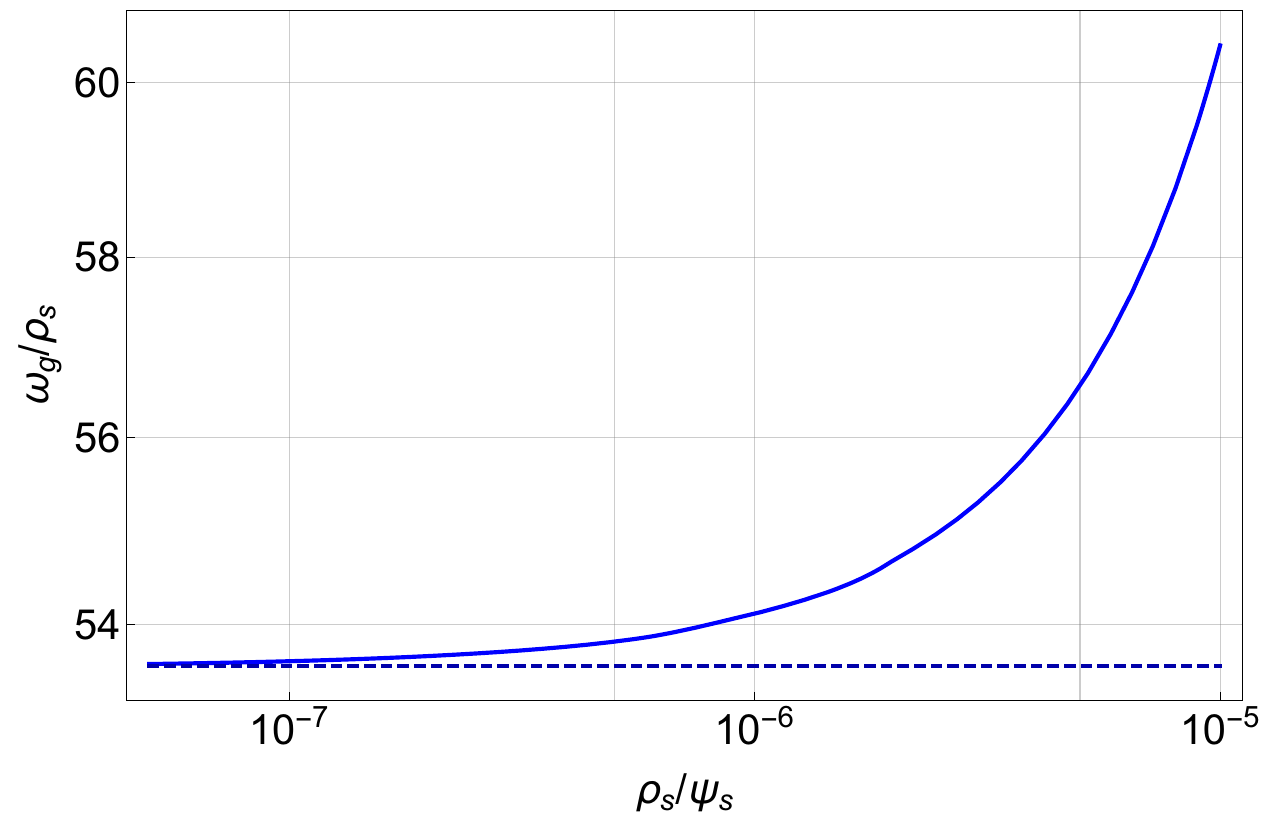}
\caption{Plot of the gap $\omega_{g}$ as a function of the pinning parameter $\delta\rho_s$.}
%Here $\psi_s=1$ and $c=1,q=-1$.}
\label{fig:gap}
\end{figure}

In Figure \ref{fig:gap}, we plot the gap as a function of the pinning parameter $\delta\rho_s$. The dashed line corresponds to the analytic result. We see that for small values of $\delta\rho_s$ the gap scales linearly with the pinning parameter $\omega_g\sim \delta\rho_s$, with the proportionality constant given precisely by $\rho_v/\sigma_b\sim 53.57$ as predicted in equation \eqref{eq:gap}. As expected, for larger values of $\delta\rho_{s}$ we observe deviations from the above formula.
 
\section{Summary and discussion}\label{sec:discussion}

In this paper we derived a hydrodynamic description of the gapless mode that emerges as a result of a holographic phase transition in which a global $U(1)$ symmetry is spontaneously broken in the bulk. In section \ref{sec:diffusion} we computed explicitly the long wavelength and small frequency retarded Green's function \eqref{eq:G_undeformed} of boundary operators which are generated by the bulk symmetry action on condensed ones. The anticipated hydrodynamic pole of \eqref{eq:G_undeformed} turned out to be diffusive, as one might have expected by the lack of involvement of a conserved charge.

Moreover, in section \ref{sec:gap} we introduced a perturbative symmetry breaking deformation which acted as a restoring force for the gapless mode of the undeformed theory and therefore introduced a gap. As a result of the deformation, the thermodynamic susceptibilities of the operators we studied in section \ref{sec:diffusion} become finite but parametrically large, reflecting the fact that we have gapped a hydrodynamic mode. By computing the retarded Green's function of this operator, we managed to give an analytic expression for the gap in \eqref{eq:gap}.

In section \ref{sec:numerics} we checked numerically the key results \eqref{eq:G_deformed} and \eqref{eq:gap} in a simple setup where the phase transition can be realised. In the context of a  CFT, thermal transitions can occur only after introducing a scale which can set the critical temperature. The way we chose to do this was by deforming the theory with a relevant operator dual to the bulk field $\psi$. The bulk Lagrangian was chosen so that it admits an IR fixed point at which the flow could terminate at low energies and which was dynamically unstable against perturbations of the complex field $Z=\phi\,e^{i\chi}$. The UV fixed point is stable and therefore our normal phase black holes are stable at high temperatures. These ingredients are not necessary but they guarantee the existence of a broken phase branch of black holes along which the field theory dual of $Z$ will take a VEV spontaneously.

In the case where translations are explicitly broken by a holographic lattice, the modes involved in thermoelectric transport are purely diffusive \cite{Hartnoll:2014lpa,Davison:2014lua,Donos:2017gej,Donos:2017ihe} resulting from the conservation of energy and charge. In this paper we described in detail the long wavelength dynamics of the global phase of a complex order parameter in the broken phase. This turned out to be diffusive as well but describing very different physics.

The complex order parameter in the black hole phases that we considered here does not break translations at thermodynamic equilibrium. However, when it does by having a phase which is linear in one of the spatial coordinates, the global phase symmetry can also be thought of as a translation of the order parameter in space. This is very similar to the physics of phasons in incommensurate crystals. There, the free energy is independent of the position of the modulation with respect to the basic structure of the crystal. The phason is a diffusive mode which is the result of this symmetry. The dispersion relation of that mode can be either linear or purely diffusive \cite{PhysRevB.25.1791,PhysRevLett.49.1833}. In this context we have identified a scenario where the phason would be diffusive; it would be interesting to construct holographic models where the dispersion relation would be linear instead of diffusive.

When the order parameter itself breaks translations, the two types of diffusive modes that describe different physics mix together in a rather non-trivial way. An interesting future direction to pursue is to study the details of the mixing of the hydrodynamic mode we considered here with the modes involved in thermoelectric transport \cite{Donos:2019hpp}, both analytically in the hydrodynamic limit and numerically by studying quasinormal modes. It is this mixing that led to the low frequency, $\omega\ll\delta\omega_g$, behaviour of the thermoelectric conductivities seen in \cite{Donos:2019tmo}, which actually resembles the conductivity measurements for pinned density waves.

Over the last few years, there have been interesting observations regarding bounds on the thermoelectric hydrodynamic modes of fully explicit holographic lattices \cite{Grozdanov:2015qia, Blake:2016sud}. However, in situations like we described in the previous paragraphs, the new collective degrees of freedom we described in our paper will mix with thermoelectric transport. It would be interesting to examine the previously proposed bounds in such a scenario.

 \section*{Acknowledgements}
 
AD, DM and CP are supported by STFC grant ST/P000371/1. VZ is supported by the China Postdoctoral Science Foundation (International Postdoctoral Fellowship Program 2018) and the National Natural Science Foundation of China (NSFC) (Grant number 11874259).
%\newpage

\bibliographystyle{utphys}
\bibliography{refs}{}

\end{document}